\documentclass[universe,article,accept,pdftex,oneauthor]{Definitions/mdpi}
\firstpage{1} 
\pubvolume{9}
\issuenum{7}
\articlenumber{318}
\pubyear{2023}
\copyrightyear{2023}
\datereceived{5 June 2023}
\daterevised{23 June 2023}
\dateaccepted{29 June 2023}
\datepublished{1 July 2023}
\hreflink{https://doi.org/10.3390/\linebreak{universe9070318}} % If needed use \linebreak
\pdfoutput=1 % Uncommented for upload to arXiv.org

\newcommand{\sqsntwo}{\mbox{$\sqrt{s_{_{\text{NN}}}}=5.02$~TeV}}
\newcommand{\sqrtsNN}{\sqrt{s_{_{\text{NN}}}}}
\newcommand{\mT}{m_{\mathrm{T}}}

\newcommand{\KT}{K_{\mathrm{T}}}
\newcommand{\TeV}{~\text{TeV}}
\newcommand{\GeV}{~\text{GeV}}
\newcommand{\MeV}{~\text{MeV}}
\Title{Centrality-Dependent Lévy HBT Analysis in $\sqrt{s_{_{\text{NN}}}}=5.02$~TeV PbPb Collisions at CMS}%Attention: title altered. Please ensure that the intended meaning has been retained. A: Fine.

\TitleCitation{Centrality-Dependent Lévy HBT Analysis in {\sqsntwo} PbPb Collisions at CMS}

\Author{Bal{\'a}zs %MDPI: Please carefully check the accuracy of names and affiliations. A: Good.
 Kórodi on behalf of the %MDPI: Since the work is presented by the author on behalf of a group/team, we revised the format, please check. A: Good.
 CMS Collaboration}

\AuthorNames{Balázs Kórodi for the CMS Collaboration}

\AuthorCitation{Kórodi, B., on behalf of the CMS Collaboration.}
\address[1]{%
Department of Atomic Physics, Faculty of Science, Eötvös Loránd University, %MDPI: For university, please add specific Department/School/Faculty/Campus. A: Done.
P\'azm\'any P\'eter s\'et\'any 1/A, H-1111 Budapest, %MDPI: Please add the postal code (or ZIP code in the U.S.). If the postal code is not available, please provide the P.O. Box. A: Done.
 Hungary; balazs.korodi@cern.ch}

\abstract{The measurement of two-particle Bose--Einstein momentum correlation functions are presented using $\sqrt{s_{_{\text{NN}}}}=5.02$~TeV PbPb collision data, recorded by the CMS experiment in 2018. The measured correlation functions are discussed in terms of Lévy-type source distributions. The Lévy source parameters are extracted as functions of transverse mass and collision centrality. These source parameters include the correlation strength $\lambda$, the Lévy stability index  $\alpha$, and the Lévy scale parameter $R$. The source shape, characterized by $\alpha$, is found to be neither Gaussian nor Cauchy. A hydrodynamic-like scaling of $R$ is also observed.}

\keyword{heavy ions; quark--gluon plasma; femtoscopy; Lévy HBT} 

\begin{document}

%%%%%%%%%%%%%%%%%%%%%%%%%%%%%%%%%%%%%%%%%%

\section{Introduction}

The investigation of the femtometer-scale space--time geometry of high-energy heavy-ion collisions has been an important area, called femtoscopy, of~high-energy physics for several decades~\cite{Lednicky:2001qv}. The~main idea of this field originates from astronomy, since it is analogous with the well-known Hanbury Brown and Twiss (HBT) effect that describes the intensity correlation of photons~\cite{HanburyBrown:1956bqd,Glauber:1962tt}. In~high-energy physics, however, the~observable is the quantum-statistical momentum correlation of hadrons, which carries information about the femtometer-scale structure of the particle-emitting source~\cite{Csorgo:1999sj,Wiedemann:1999qn}. The~measurements of such momentum correlations are partially responsible for establishing the fluid nature of the quark--gluon plasma (QGP) created in heavy-ion collisions~\cite{Adler:2004rq,Csorgo:1995bi}. Furthermore, the~measured source radii provide information about the transition from the QGP to the hadronic phase~\mbox{\cite{Csanad:2004mm,Pratt:2008qv}}, as well as about the phase space of quantum chromodynamics~\cite{Lacey:2014wqa}. 

Recent high-precision femtoscopic measurements~\cite{PHENIX:2017ino,NA61SHINE:2023qzr} have shown that the previously widely assumed Gaussian~\cite{Adler:2004rq,STAR:2004qya,ALICE:2015hvw} or Cauchy~\cite{CMS:2017mdg,ATLAS:2017shk} source distributions do not provide an adequate description of the measured correlation functions. Instead, a~generalization of these distribution, the~Lévy alpha-stable distribution~\cite{UchaikinZolotarev+2011}, is needed for a statistically acceptable description~\cite{PHENIX:2017ino,NA61SHINE:2023qzr}. The~shape of the Lévy distribution is characterized by the Lévy stability index $\alpha$, and~can be influenced by various physical phenomena, e.g.,~anomalous diffusion~\cite{Metzler:1999zz,Csorgo:2003uv, Csanad:2007fr}, resonance decays~\cite{Kincses:2022eqq,Korodi:2022ohn}, jet fragmentation~\cite{Csorgo:2004sr}, and critical phenomena~\cite{Csorgo:2005it}. Until~now, the~$\alpha$ parameter had not been measured at the largest energies accessible at the LHC. The~question of how $\alpha$ changes compared to lower energies signifies the need for a Lévy HBT analysis at LHC~energy.

In this paper, the~Lévy HBT analysis of two-particle Bose--Einstein momentum correlations is presented using {\sqsntwo} PbPb collision data recorded by the CMS experiment. The~source parameters, extracted from the correlations functions, are studied as functions of transverse mass and collision centrality. 
%%%%%%%%%%%%%%%%%%%%%%%%%%%%%%%%%%%%%%%%%%
\section{Femtoscopy with Lévy~Sources}

The quantum-statistical momentum correlation of identical bosons is called Bose--Einstein correlation. This correlation is in connection with the source function $S(x,p)$~\cite{Csorgo:1999sj,Wiedemann:1999qn}, which is the phase-space probability density of particle production at space--time point $x$ and four-momentum $p$. After~some approximations detailed in Refs.~\cite{Csorgo:1999sj,Wiedemann:1999qn}, the following formula is obtained:
\begin{equation} \label{e:Cqdef}
    C^{(0)}(Q,K)\approx 1+\frac{|\widetilde{S}(Q,K)|^2}{|\widetilde{S}(0,K)|^2},
\end{equation}
where $C^{(0)}(Q,K)$ is the two-particle momentum correlation function, $Q$ is the pair relative four-momentum, $K$ is the pair average four-momentum, the~superscript $(0)$ denotes the neglection of final-state interactions, and~$\widetilde{S}(Q,K)$ is the Fourier transform of the source with
\begin{equation}
    \widetilde S(Q,K)=\int S(x,K) e^{iQx} d^4x.
\end{equation}

Equation~(\ref{e:Cqdef}) implies that $C^{(0)}(Q=0,K)=2$. In~previous measurements, it was found, however, that $C^{(0)}(Q\rightarrow0,K)<2$. This result can be understood via the core--halo model~\cite{Bolz:1992hc,Csorgo:1994in}, wherein the source is divided into two parts, a~core of primordial hadrons and a halo of long-lived resonances. The~halo is experimentally unresolvable due to its large size, which leads to small momentum in Fourier space. If~$S$ represents only the core part of the source, its connection to the correlation function becomes
\begin{equation} \label{e:Cqlambda}
    C^{(0)}(Q,K)\approx 1+\lambda\frac{|\widetilde{S}(Q,K)|^2}{|\widetilde{S}(0,K)|^2},
\end{equation}
where $\lambda$ is the square of the core fraction, and~it is often called the correlation strength~parameter.

Using Equation~(\ref{e:Cqlambda}), a~theoretical formula for $C^{(0)}(Q,K)$ can be calculated by assuming a given source distribution. In~this analysis, a~generalization of the Gaussian distribution, the~so-called spherically symmetric Lévy alpha-stable distribution~\cite{UchaikinZolotarev+2011}, was assumed for the spatial part of the source. This distribution is defined by the following Fourier transform in three dimensions: %MDPI: Please confirm if the bold format in formulas should be retained. Please check all bold equations in the tex. A: They should be retained everywhere.
\begin{equation}
    \mathcal{L}(\boldsymbol{r};\alpha,R)=\frac{1}{(2\pi)^3}\int\mathrm{d^3}\boldsymbol{q}\, e^{i\boldsymbol{qr}} e^{-\frac{1}{2}|\boldsymbol{q}R|^{\alpha}},
\end{equation}
where $\boldsymbol{q}$ is an integration variable, $\boldsymbol{r}$ is the variable of the distribution, $\alpha$ and $R$ are parameters; the~Lévy stability index and the Lévy scale parameter, respectively. The~$\alpha$ parameter describes the shape of the distribution, with~$\alpha=2$ corresponding to the Gaussian and $\alpha=1$ to the Cauchy case. The~$R$ parameter describes the spatial scale of the source, as~it is proportional to the full width at half maximum. There are many possible reasons~\cite{Metzler:1999zz,Csorgo:2003uv,Csorgo:2004sr,Csorgo:2005it,Csanad:2007fr,Kincses:2022eqq,Korodi:2022ohn} behind the appearance of the Lévy distribution in heavy-ion collisions, but~these possibilities are still under investigation by the community. In~case of a spherically symmetric Lévy source, the~two-particle correlation function has the form~\cite{Csorgo:2003uv}
\begin{equation}
    C^{(0)}(q)=1+\lambda e^{-(qR)^{\alpha}},
\end{equation}
where $q=|\boldsymbol{Q}|$ is the magnitude of the spatial part of $Q$.

In the above formulas, the~presence of final-state interactions was neglected. In~the case of charged particles, the~most important final-state interaction is the Coulomb interaction, which is usually taken into account in the form of a Coulomb correction%A: Added 3 references for the Coulomb correction.
$K_C(q;R,\alpha)$~\cite{Kurgyis:2020vbz,Kincses:2019rug,Csanad:2019cns}. Using the Bowler--Sinyukov method~\cite{Sinyukov:1998fc}, one obtains
\begin{equation}\label{e:C2:Coulomb}
C(q)= 1-\lambda  + \lambda  (1+e^{-(qR)^{\alpha}})K_{C}(q;R,\alpha).
\end{equation}
In %MDPI: Please check if all no indentation for upper-case below Equations is necessary or not. A: I checked all of them and removed the identation where I thought it is not needed. Feel free to put them back in case it violates some journal specific rule.
this analysis, the~$R$ and $\alpha$-dependent Coulomb correction, calculated in Ref.~\cite{Csanad:2019lkp}, was utilized. A~formula based on Equation~(\ref{e:C2:Coulomb}) was used for fitting to the measured correlation~functions.

\section{Measurement~Details}

The used data sample contains $4.27\times 10^{9}$ PbPb events at a center-of-mass energy per nucleon pair of {\sqsntwo}, recorded by the CMS experiment in 2018. The~detailed description of the CMS detector system can be found in Ref.~\cite{CMS:2008xjf}. For~the analysis, only events with precisely one nucleus--nucleus collision were used, where the longitudinal distance of the interaction point from the center of the detector was also less than 15 cm. Further event selections were applied to reject events from beam--gas interactions and nonhadronic collisions~\cite{CMS:2016xef}. The~individual tracks were filtered based on their transverse momentum, pseudorapidity, distance to the vertex, the~goodness of the track fit, and~the number of hits in the tracking~detectors. 

Particle identification in central PbPb collisions is not possible with the CMS detector; therefore, all charged tracks passing the other selection criteria were used. The~majority of these charged particles are pions~\cite{ALICE:2019hno}, so the pion mass was assumed for all of them. The~largest contamination is caused by kaons and protons~\cite{ALICE:2019hno}, and~this effect is discussed in Section~\ref{s:results}.

\textls[-5]{Measuring two-particle Bose--Einstein correlation functions means measuring pair distributions. Besides~the quantum-statistical effects, these pair distributions are influenced by detector acceptance, kinematics, and~other phenomena. In~order to remove these unwanted effects, the~correlation function is calculated as the normalized ratio of two distributions, the~actual (signal) distribution $A(q)$, and the background distribution $B(q)$, with}
\begin{equation} \label{e:ABq}
    C(q)=\frac{A(q)}{B(q)}\frac{\int B(q) dq}{\int A(q) dq},
\end{equation}
where the integrals are calculated over a range where the quantum-statistical effects are not present. The~$A(q)$ distribution contains all same charged pairs of a given event, while the $B(q)$ distribution contains all same charged pairs of a mixed event. This mixed event is obtained by randomly selecting particles from different events, as~detailed in Refs.~\cite{Achard:2011zza,PHENIX:2017ino}. For~the validity of Equation~(\ref{e:ABq}), it was assumed that the produced particles had a uniform rapidity distribution~\cite{Vechernin:2019dki}.

In the measurement of $C(q)$, the~$q$ variable is taken as the magnitude of the relative momentum in the longitudinally comoving system (LCMS), where the longitudinal component of the average momentum is zero. This coordinate system was chosen because, in earlier measurements, the source was found to be approximately spherically symmetric in this frame~\cite{Adler:2004rq}. The~measurement is carried out up to $q=8$ GeV/$c$ in 6 centrality (0--60\%) and 24 average transverse momentum $\KT$ (0.5--1.9~GeV/$c$) classes, separately for positively and negatively charged pairs. In~order to remove the merging and splitting effects caused by the finite resolution of the tracking detectors, a~pair selection was applied. These artifacts were limited to a region with small $\Delta \eta$ and $\Delta \phi$; therefore, each pair had to satisfy the following condition:
\begin{equation}
    \left ( \frac{|\Delta \eta|}{0.014} \right )^2 + \left ( \frac{|\Delta \phi|}{0.022} \right )^2 >1,
\end{equation}
where $\Delta \eta$ is the pseudorapidity difference and $\Delta \phi$ is the azimuthal angle difference. Tracking efficiency correction factors were also utilized when measuring the $A(q)$ and $B(q)$ distributions.

Even after removing most of the non-quantum-statistical effects by taking the ratio of $A(q)$ and $B(q)$, a~structure was observed in $C(q)$ at large $q$ values, where the quantum-statistical effects were not present. This long-range background can be the result of phenomena such as energy and momentum conservation, resonance decays, bulk flow~\cite{CMS:2017mdg}, and~minijets~\cite{CMS:2017mdg}. To~remove any potential influence of the long-range background on the low $q$ region where the Bose--Einstein peak is present, $C(q)$ was divided by a background function $BG(q)$, resulting in the double-ratio correlation function $DR(q)$:
\begin{equation}
    DR(q)=\frac{C(q)}{BG(q)}.
\end{equation}
The explicit form of $BG(q)$ was determined by fitting the following empirically determined formula~\cite{CMS:2019fur,CMS:2017mdg,CMS:2023jjt} to the large $q$ part of $C(q)$:
\begin{equation}
    BG(q)=N \left( 1+\alpha_1 e^{-(qR_1)^2} \right) \left( 1-\alpha_2 e^{-(qR_2)^2} \right),
\end{equation}
where $N, \alpha _1, \alpha _2, R_1, R_2$ are fit parameters with no physical~meaning.

\begin{figure}[H]
%\centering
\includegraphics[scale=0.4]{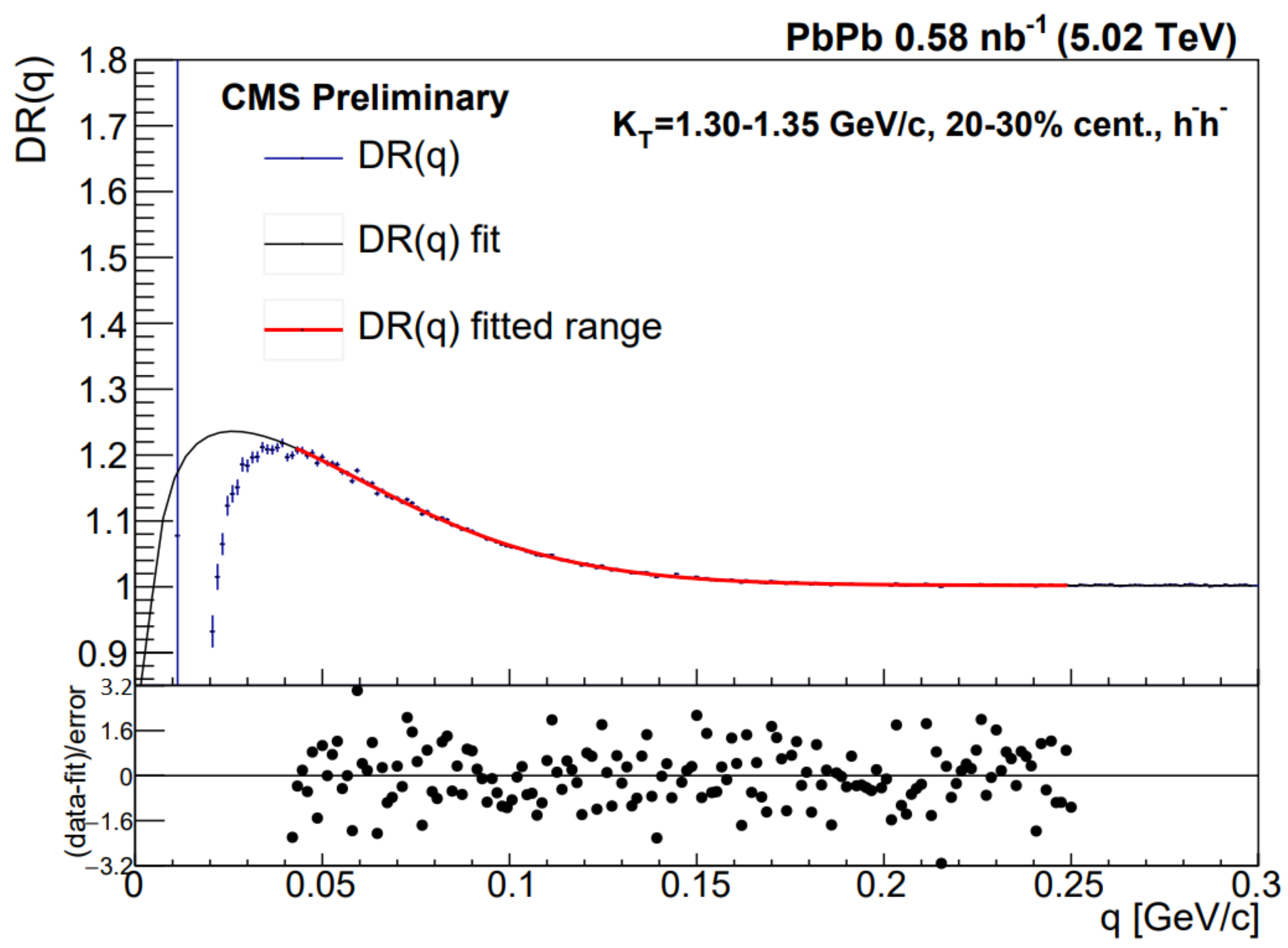}
\caption{An
%MDPI: 1. Please change the hyphen (-) into minus sign ($-$, "U+2212"). e.g., "-1" should be "$-$1". A: Unfortunately, the codes that were used to create all these Preliminary figures are not available, I only have older and newer versions of them. Since the only requested modification is an extremely minor one, I would like to ask if it is possible to publish these figures as they are now? If you insist to have the minus signs written correctly, I can either use a pdf editor, where the end result is not guaranteed to look better, or I would need to rewrite the codes exactly as they were, which would take a huge amount of time.  
%MDPI: 2. Number at the left of the image is incomplete, please revise. A: Done using a pdf editor.
 example fit to the double-ratio correlation function $DR(q)$ of negatively charged hadrons~\cite{PAS}. The~fitted function is shown in black, while the red overlay indicates the range used for the fit. The~$\KT$ and centrality class is shown in the legend. The~lower panel indicates the deviation of the fit from the~data.}
\label{f:DRqfit}
\end{figure}

The $DR(q)$ distributions were fitted with the following formula based on Equation~(\ref{e:C2:Coulomb}):
\begin{equation}
    DR(q) = N (1+\epsilon q) \left[ 1-\lambda  + \lambda  (1+e^{-(qR)^{\alpha}}) K_{C}(q;R,\alpha) \right],
\end{equation}
where $N$ is a normalization parameter and a possible residual linear background is allowed through the $\epsilon$ parameter. The~fits were performed using the MINUIT2 package~\cite{James:1975dr,Brun:1997pa} and the statistical uncertainties were calculated with the MINOS algorithm~\cite{James:1975dr,Brun:1997pa}. The~lower and upper fit limits were determined individually in each centrality and $\KT$ class by selecting the limits resulting in the best fit. The~goodness of fit was measured by the confidence level, calculated from the $\chi^2$ and the number of degrees of freedom of the fit. This confidence level was in the statistically acceptable range ($>$0.1\%) for each fit. An~example fit is shown in Figure~\ref{f:DRqfit}. In~the region below approximately $q=0.05$ GeV/$c$, the~measured data are not reliable due to the finite momentum resolution and pair reconstruction efficiency of the detectors; consequently, that region was not used for~fitting.

\vspace{-3pt} 

The systematic uncertainties of $R, \alpha$, and $\lambda$ were determined by individually changing each of the analysis settings to slightly larger and smaller values, and conducting the whole analysis procedure again. The~deviations from the nominal results were then added in quadrature, resulting in the full systematic uncertainty. The~considered analysis settings were the centrality calibration, the~vertex selection, the~different track selection criteria, the~pair selection, and~the fit limits. Out of these, the~dominant sources of systematic uncertainty were the fit limits. The~full systematic uncertainty was separated into correlated and uncorrelated parts, so that the latter could be taken into account when fitting to the~parameters.

\section{Results and~Discussion} \label{s:results}

\textls[-11]{As mentioned before, the~parameters $\alpha, R$, and $\lambda$ were measured separately for positively and negatively charged hadron pairs. As~not much difference was observed between the two cases, some of the results for negatively charged pairs are shown only in Appendix~\ref{a:negativepairs}. }

The measurement was carried out in $\KT$ classes, but~in order to facilitate the comparison with previous measurements and with theory, the~parameters are presented as functions of the transverse mass $\mT$, defined as
\begin{equation}
    \mT=\sqrt{\frac{\KT^2}{c^2}+m^2},
\end{equation}
where $m$ is the mass of the investigated particle species. Although~all charged tracks were used in the analysis, the~pion mass was used for $m$, since above 90\% of the identical particle pairs were pion~pairs.

\begin{figure}[H]
%\centering
\includegraphics[scale=0.56]{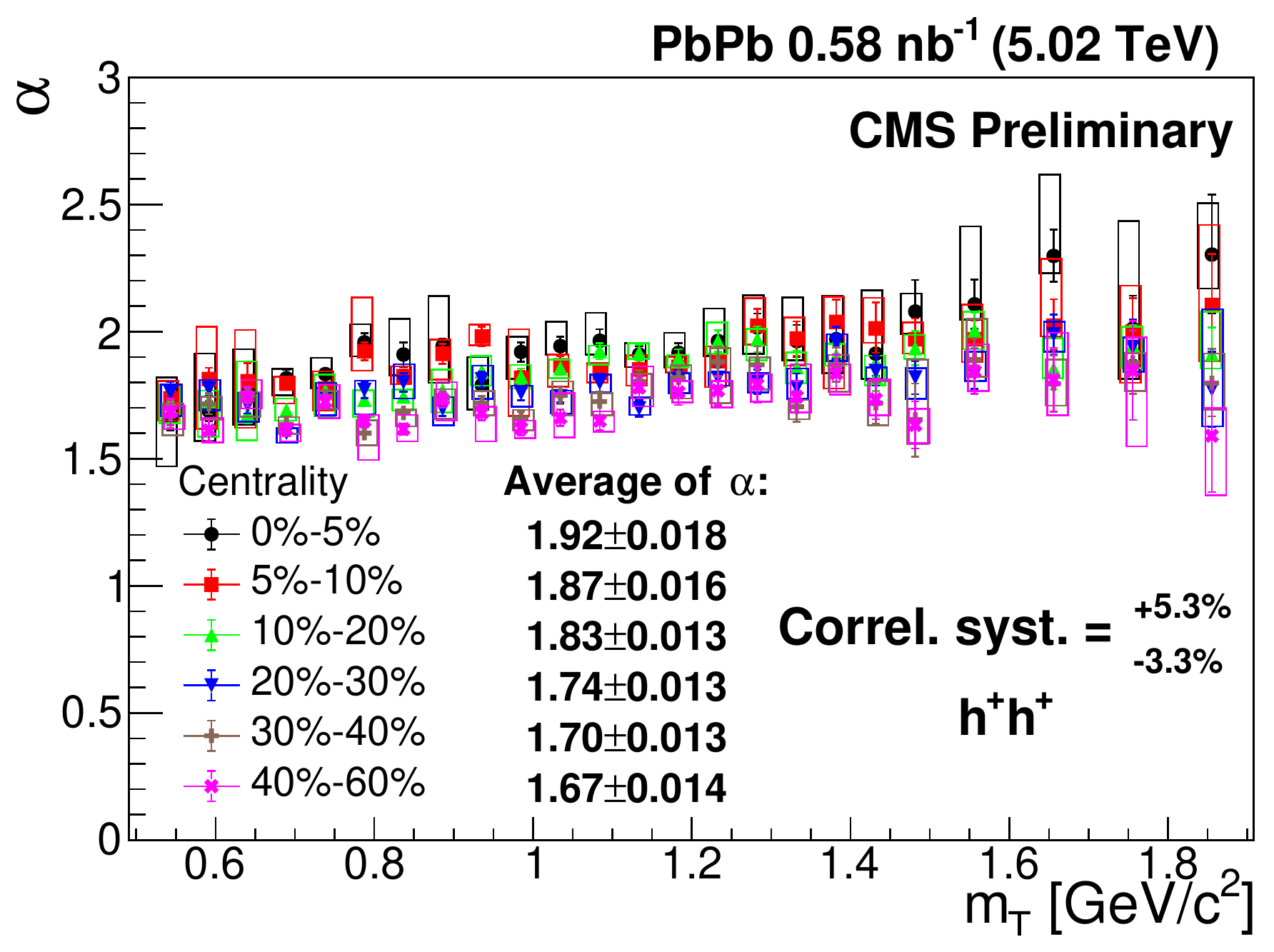}
\caption{\textls[-24]{The %MDPI: Please change the hyphen (-) into minus sign ($-$, "U+2212"). e.g., "-1" should be "$-$1".
 Lévy stability index $\alpha$ versus the transverse mass $\mT$ in different centrality classes for positively charged hadron pairs~\cite{PAS}. The~error bars are the statistical uncertainties, while the boxes indicate the uncorrelated systematic uncertainties. The~correlated systematic uncertainty is shown in the~legend.}}
\label{f:alpha}
\end{figure}

\begin{figure}[H]
%\centering
\includegraphics[scale=0.56]{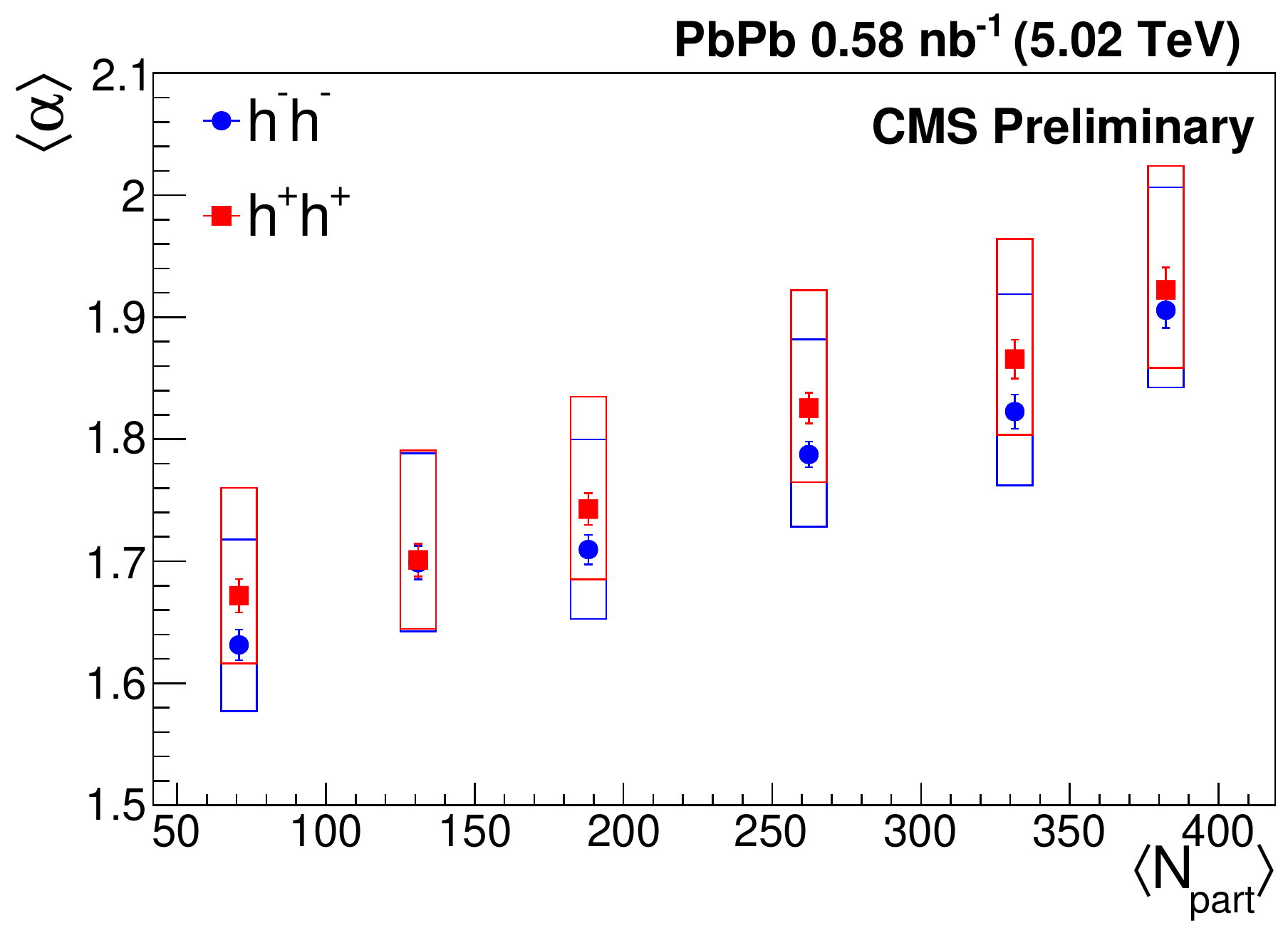}
\caption{The %MDPI: Please change the hyphen (-) into minus sign ($-$, "U+2212"). e.g., "-1" should be "$-$1".
 average Lévy stability index $\langle \alpha \rangle$ versus $\langle N_{\text{part}}\rangle$ in different centrality classes for positively and negatively charged hadron pairs~\cite{PAS}. The~error bars are the statistical uncertainties, while the boxes indicate the systematic~uncertainties.}
\label{f:average_alpha}
\end{figure}

\textls[-11]{The measured $\alpha$ values are shown in Figure~\ref{f:alpha} as a function of $\mT$, for~positively charged pairs. Within~uncertainties, most of the values are between 1.6 and 2.0, meaning that the source follows the general Lévy distribution, instead of the Gaussian. However, the~deviation from the Gaussian case is not as large as it was found for 0--30\% centrality AuAu collisions at $\sqrtsNN = 200$~GeV~\cite{PHENIX:2017ino}, where a mean value for $\alpha$ of 1.207 was obtained for pion pairs with $|\eta| < 0.35$ and $228 < \mT < 871$~MeV/$c^2$. For~a given centrality class, $\alpha$ is almost constant with $\mT$. The~average of $\alpha$ ($\langle \alpha \rangle$) is indicated in Figure~\ref{f:alpha} for each centrality class, and~it is shown in Figure~\ref{f:average_alpha} as a function of the average number of participating nucleons in the collision ($\langle N_{\text{part}}\rangle$), for~both positively and negatively charged pairs. The~$\langle N_{\text{part}}\rangle$ values were calculated for each centrality class~\cite{Loizides:2017ack}, with~a larger value corresponding to a more central case. The~$\langle \alpha \rangle$ values show a monotonic increasing trend with $\langle N_{\text{part}}\rangle$, which means that the shape of the source is $\langle N_{\text{part}}\rangle$ (or equivalently, centrality) -dependent. The~shape is closer to the Gaussian distribution in case of more central events. The~$\langle \alpha \rangle$ values are slightly higher for positively charged pairs, although~the deviations are within systematic~uncertainties. }

\vspace{-8pt}

\begin{figure}[H]
%\centering
\includegraphics[scale=0.57]{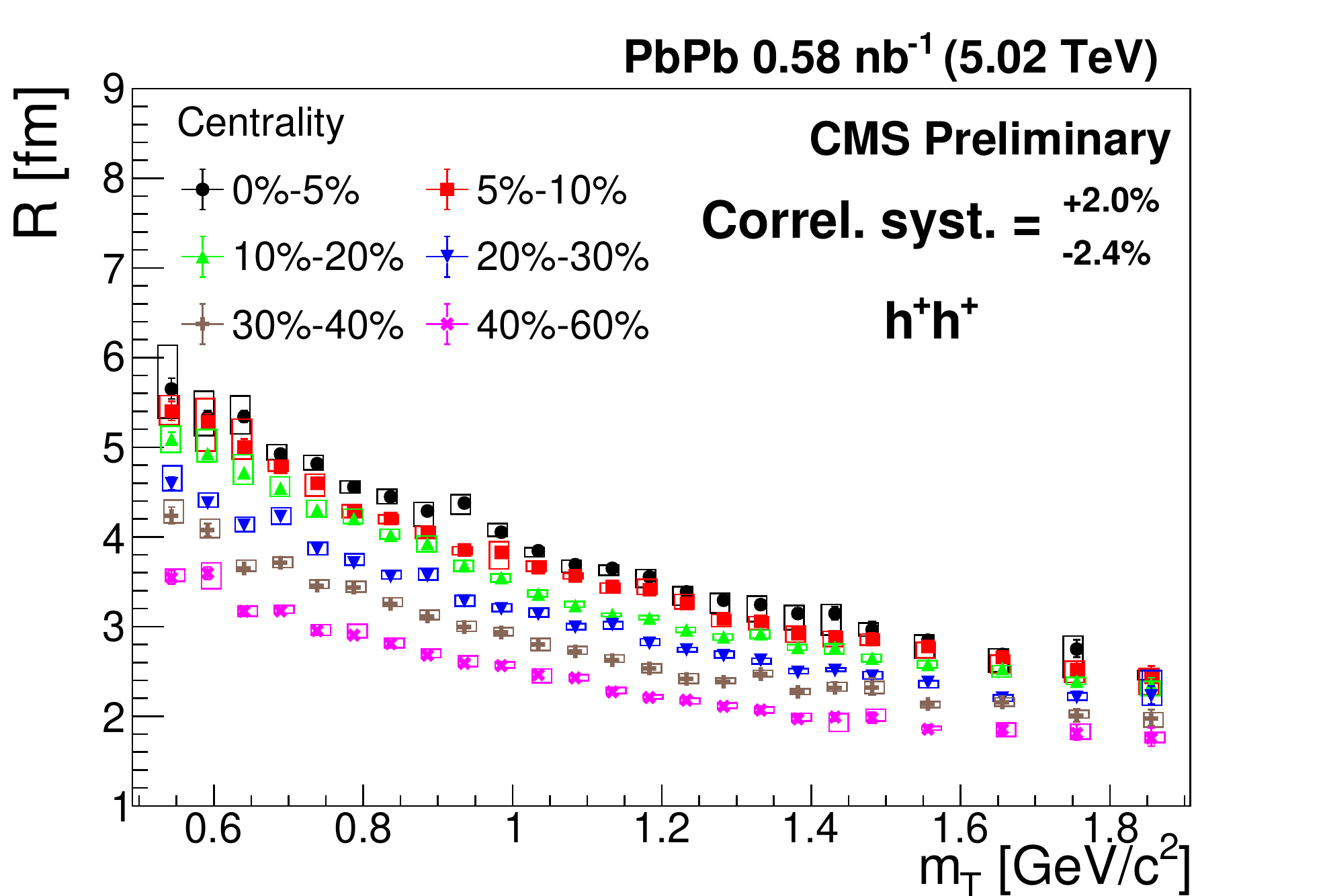}
\caption{The %MDPI: Please change the hyphen (-) into minus sign ($-$, "U+2212"). e.g., "-1" should be "$-$1".
 Lévy scale parameter $R$ versus $\mT$ in different centrality classes for positively charged hadron pairs~\cite{PAS}. The~error bars are the statistical uncertainties, while the boxes indicate the uncorrelated systematic uncertainties. The~correlated systematic uncertainty is shown in the~legend.}
\label{f:R}
\end{figure}

\begin{figure}[H]
%\centering
\includegraphics[scale=0.57]{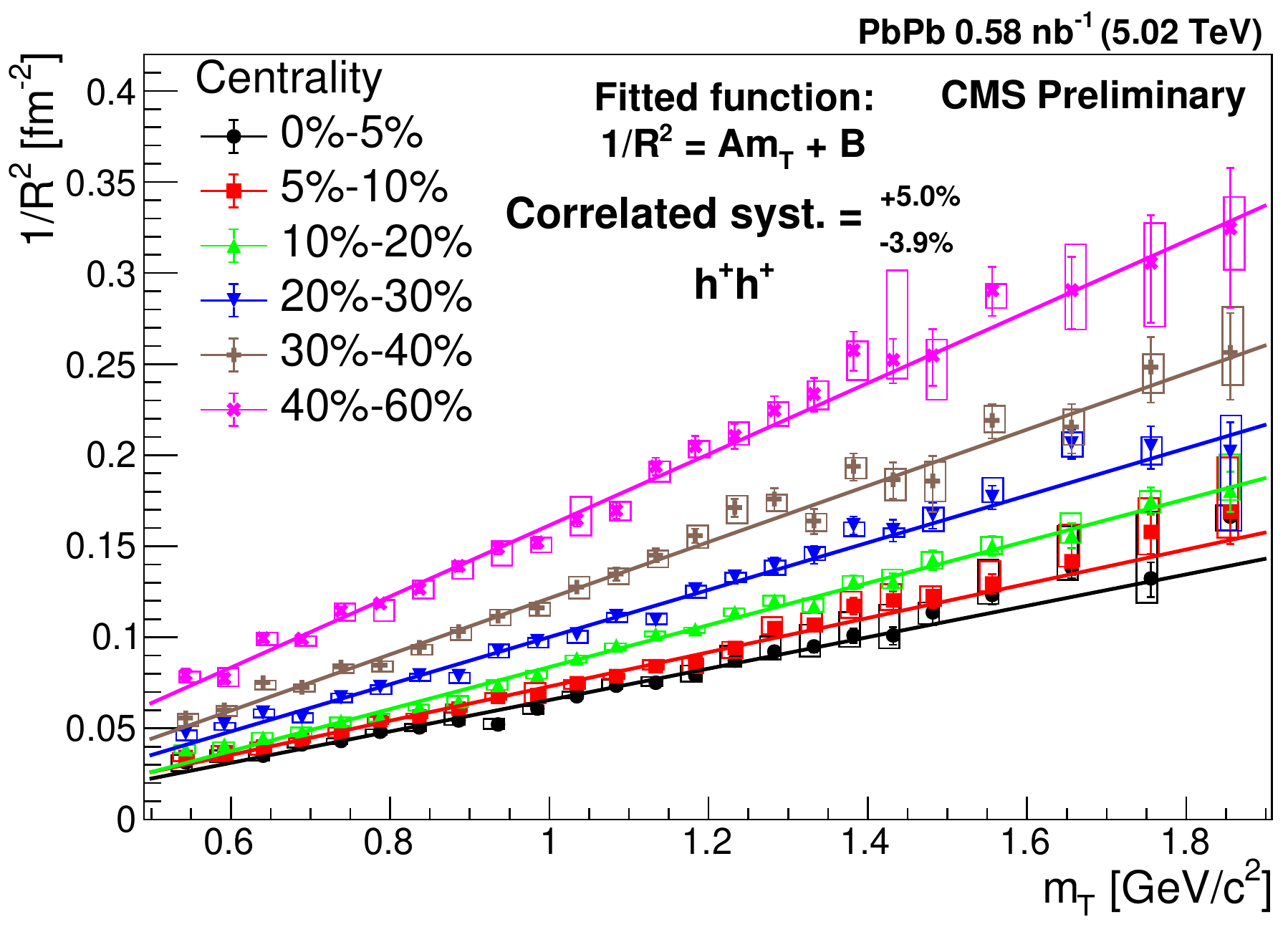}
\caption{The %MDPI: Please change the hyphen (-) into minus sign ($-$, "U+2212"). e.g., "-1" should be "$-$1".
 inverse square of the Lévy scale parameter $R$ versus $\mT$ in different centrality classes for positively charged hadron pairs~\cite{PAS}. The~error bars are the statistical uncertainties, while the boxes indicate the uncorrelated systematic uncertainties. The~correlated systematic uncertainty is shown in the legend. A~line is fitted to the data for each~centrality.}
\label{f:1_R2}
\end{figure}

The measured $R$ values are shown in Figure~\ref{f:R} as a function of $\mT$ for positively charged pairs. A~decreasing trend with $\mT$ and as the collisions become more peripheral is observed, with~the values ranging between 1.6 and 5.8~fm. The~centrality dependence confirms the geometrical interpretation of the $R$ parameter, because~a smaller source size is expected in case of more peripheral collisions. To~further investigate the $\mT$ dependence of $R$, $1/R^2$ was plotted as a function of $\mT$, as~shown in Figure~\ref{f:1_R2}. In~case of a Gaussian source, hydrodynamic models~\cite{Csorgo:1995bi,Makhlin:1987gm} predict the linear scaling
\begin{equation} \label{e:hydrofit}
    \frac{1}{R^2}=A\mT +B,
\end{equation}
where $A$ and $B$ are parameters with physical meaning. The~slope $A$ is connected to the Hubble constant ($H$) of the QGP with~\cite{Csorgo:1995bi,Chojnacki:2004ec}
\begin{equation}
    A=\frac{H^2}{T_{\mathrm{f}}},
\end{equation}
where $T_{\mathrm{f}}$ is the freeze-out temperature. The~intercept $B$ is connected to the size of the source ($R_{\mathrm{f}}$) at freeze-out with~\cite{Csorgo:1995bi,Chojnacki:2004ec}
\begin{equation} \label{e:Rf}
    B=\frac{1}{R_{\mathrm{f}}^2}.
\end{equation}

\begin{figure}[H]
%\centering
\includegraphics[scale=0.54]{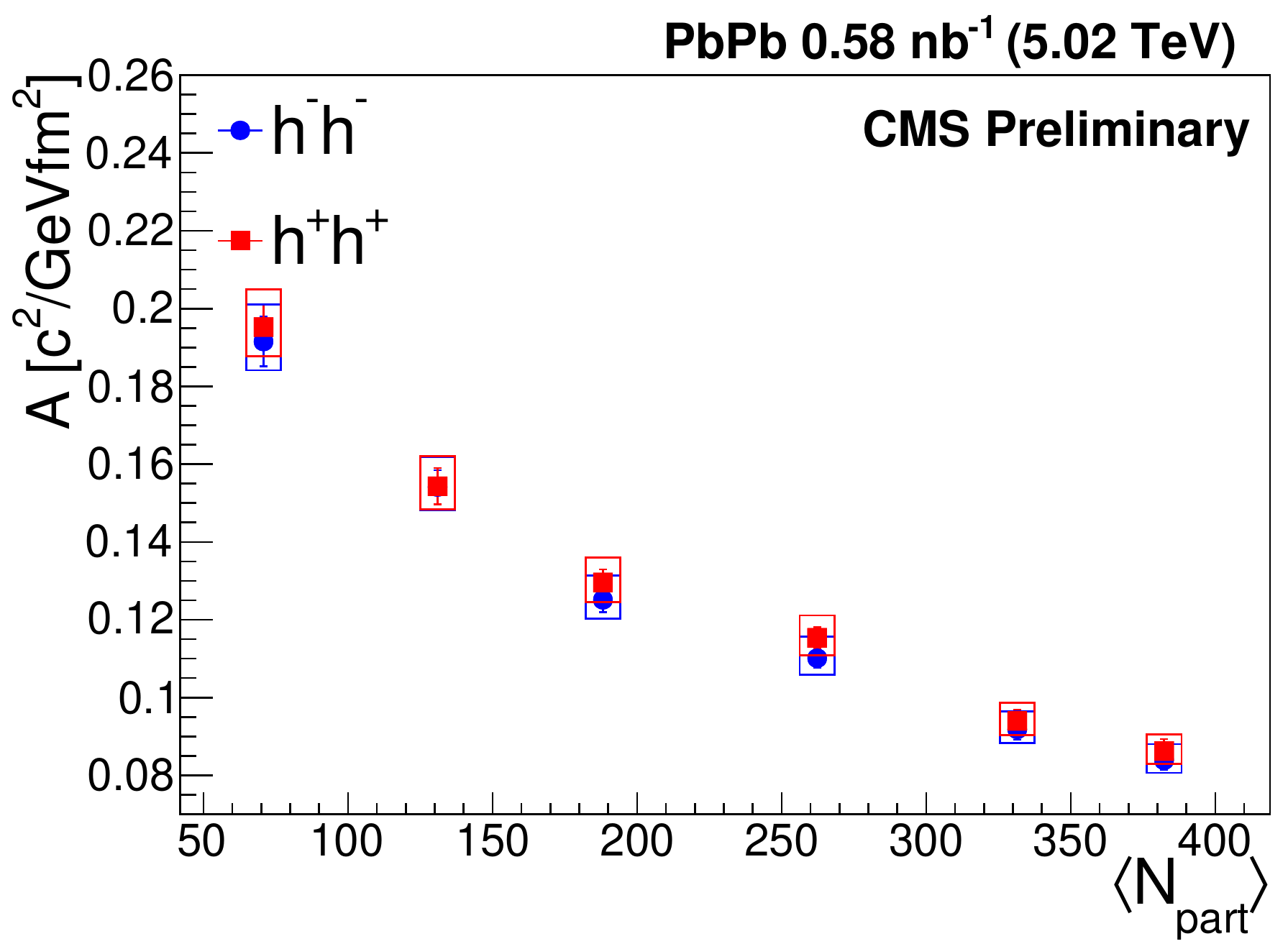}\\
\includegraphics[scale=0.534]{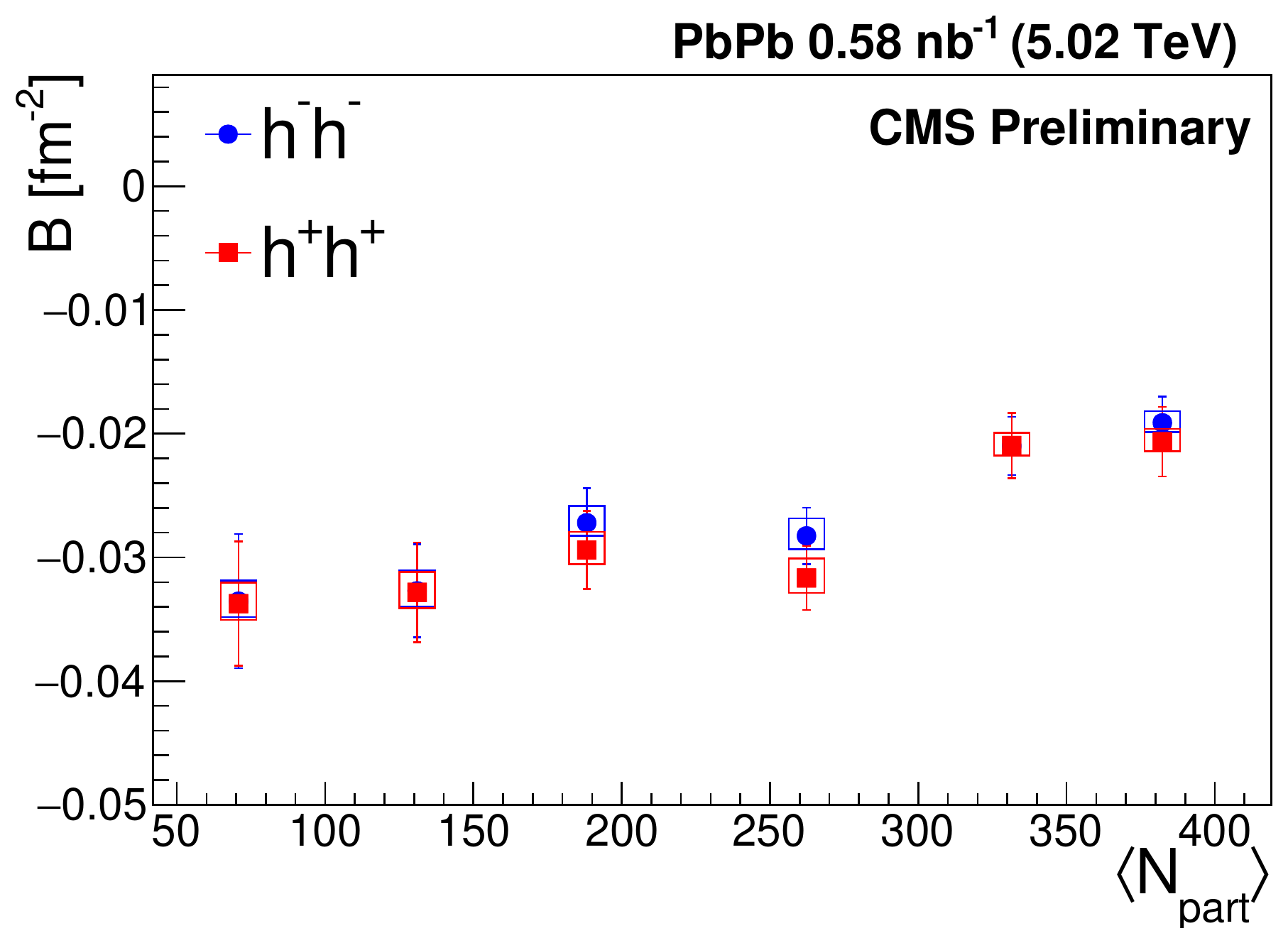}
\caption{The %MDPI: Please change the hyphen (-) into minus sign ($-$, "U+2212"). e.g., "-1" should be "$-$1".
 two fit parameters from the linear fit: the slope $A$ (\textbf{upper}) and the intercept $B$
(\textbf{lower}) versus $\langle N_{\text{part}}\rangle$ for negatively and positively charged hadron pairs~\cite{PAS}. The~error bars are the statistical
uncertainties, while the boxes indicate the systematic uncertainties.}
\label{f:ABvsNpart}
\end{figure}

In order to verify whether the linear scaling also holds in the Lévy case, a~linear fit was performed for each centrality class using Equation~(\ref{e:hydrofit}). The~statistical uncertainty and the uncorrelated systematic uncertainty of $1/R^2$ was added in quadrature and~used for determining the $\chi^2$ of the fits. In~this way, the~confidence levels were statistically acceptable for each centrality class, showing that a hydrodynamic-like scaling holds for a Lévy source as well. The~fitted lines are shown in Figure~\ref{f:1_R2}, and~the fit parameters ($A$ and $B$) are shown in Figure~\ref{f:ABvsNpart} as functions of $\langle N_{\text{part}}\rangle$, for~both positively and negatively charged pairs. By~assuming a constant freeze-out temperature of $T_{\mathrm{f}}=156$~MeV~\cite{Steinbrecher:2018phh}, the~Hubble constant falls between 0.12~$c$/fm and 0.18~$c$/fm. Due to the fact that~the $A$ parameter decreases toward more central collisions (larger $\langle N_{\text{part}}\rangle$), the~Hubble constant also decreases, making the speed of the expansion lower in central collisions. The~$B$ parameter has a negative value in each case, which makes it impossible to calculate a freeze-out size using Equation~(\ref{e:Rf}). The~reasons behind a negative intercept and~the interpretation of this result are currently unknown. This may be connected to fluctuations in the initial state~\cite{Sputowska:2019yvr} which were not taken into account in the hydrodynamic~models.

\vspace{-4pt} 
\vspace{-8pt}

%\vspace{-8pt}

\begin{figure}[H]
%\centering
\includegraphics[scale=0.54]{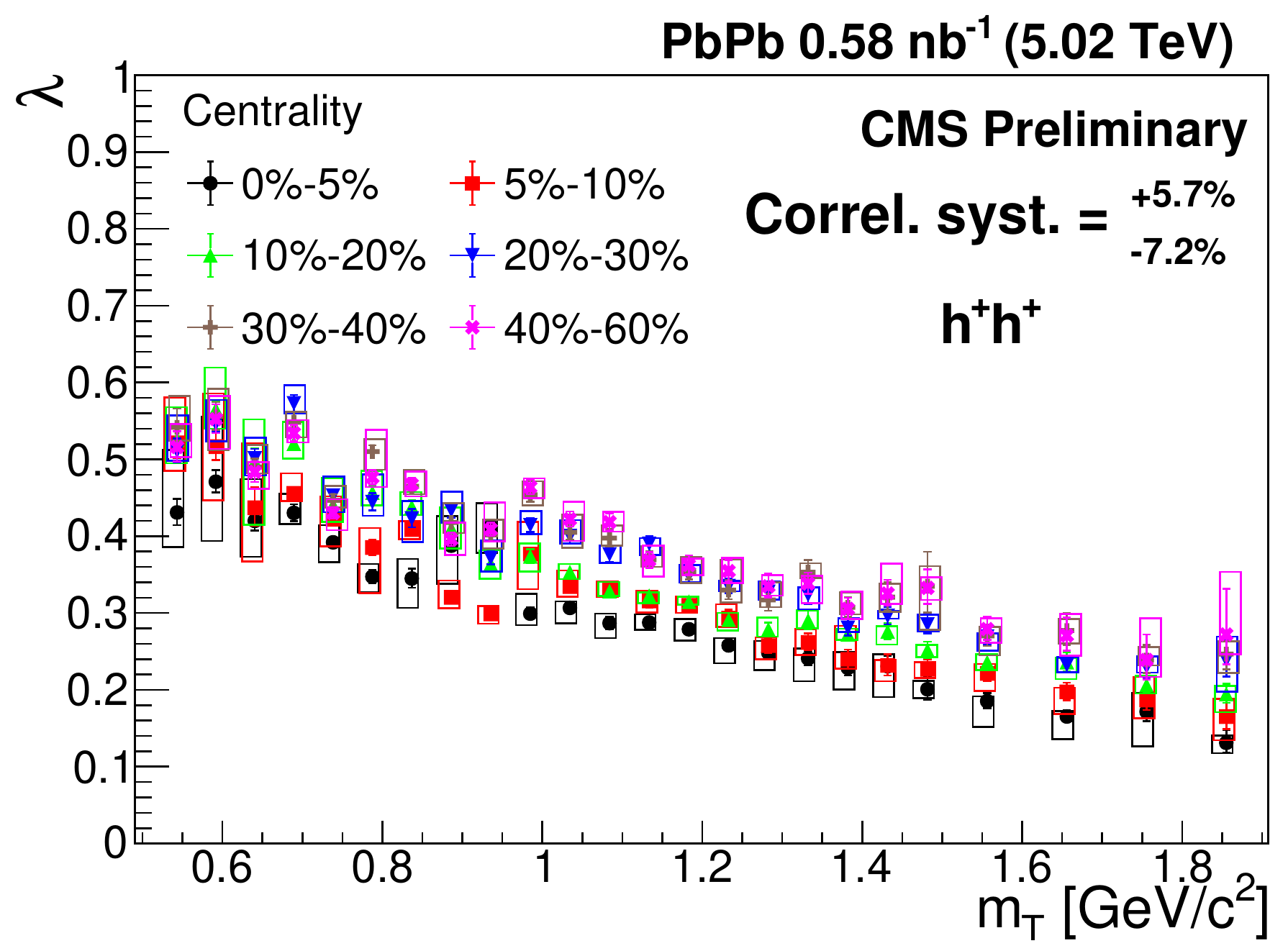}\\
\includegraphics[scale=0.54]{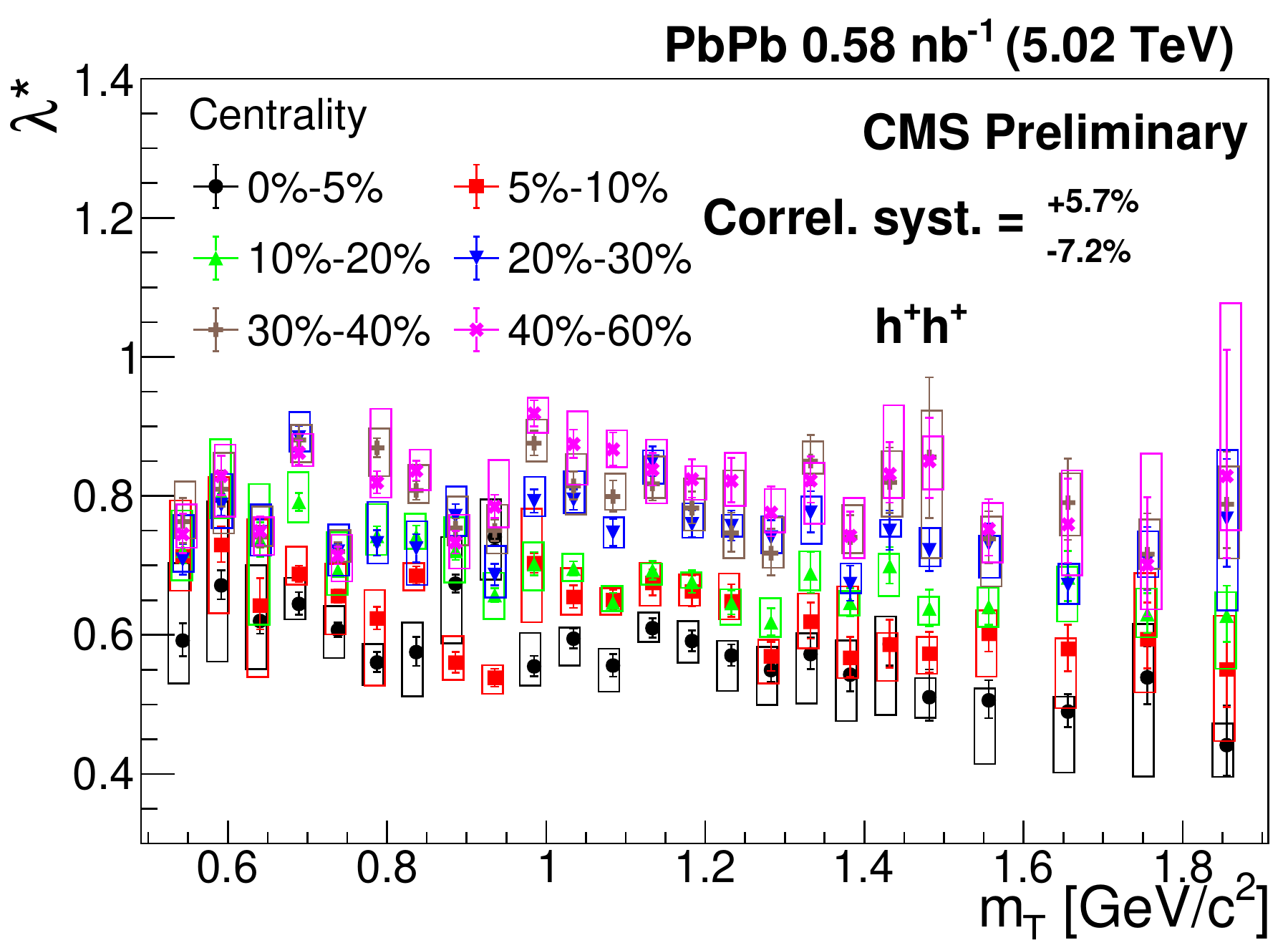}
\caption{The %MDPI: Please change the hyphen (-) into minus sign ($-$, "U+2212"). e.g., "-1" should be "$-$1".
 correlation strength $\lambda$ and the rescaled correlation strength $\lambda^*$ versus $\mT$ in different centrality classes for positively charged hadron pairs~\cite{PAS}. The~error bars are the statistical uncertainties, while the boxes indicate the uncorrelated systematic uncertainties. The~correlated systematic uncertainty is shown in the~legend.}
\label{f:lambda}
\end{figure}

The measured $\lambda$ values are shown in the upper panel of Figure~\ref{f:lambda} as a function of $\mT$, for~positively charged pairs. A~decreasing trend with $\mT$ as the collisions became more central is observed. In~case of identified particles, $\lambda$ is the square of the ratio of core particles. Due to the lack of particle identification, our sample contained particles other than pions, mostly kaons and protons. As~a result of this contamination, $\lambda$ was suppressed by a factor of the square of the pion fraction. The~pion fraction was measured by the ALICE Collaboration~\cite{ALICE:2019hno}, and~it decreased with $\mT$, resulting in the decreasing trend of $\lambda$ in the upper panel of Figure~\ref{f:lambda}. For~the $\alpha$ and the $R$ parameters, a~characteristic $\mT$ dependence was observed; thus, these parameters could not have been influenced by the $\mT$-dependent effect of the lack of particle identification. To~remove the effect of the contamination from $\lambda$, the~$\lambda^*$ parameter was introduced by rescaling $\lambda$ with the square of the pion fraction:
\begin{equation}
    \lambda^*=\frac{\lambda}{(N_{\text{pion}}/N_{\text{hadron}})^2}.
\end{equation}

The rescaled correlation strength $\lambda^*$ is shown in the lower panel of Figure~\ref{f:lambda}. Compared to $\lambda$, the~decreasing trend with $\mT$ is no longer shown in the data, suggesting that it was caused purely by the lack of particle identification. The~centrality dependence, on~the other hand, remained the same, which means that the fraction of core pions is smaller in more central~collisions.

%%%%%%%%%%%%%%%%%%%%%%%%%%%%%%%%%%%%%%%%%%
\section{Conclusions}

In this paper, a~centrality-dependent Lévy HBT analysis of two-particle Bose--Einstein correlations was presented, using {\sqsntwo} PbPb collision data recorded by the CMS experiment. The~measured correlation functions were described by the assumption of a Lévy alpha-stable source distribution.
Three source parameters, the~Lévy stability index $\alpha$, the~Lévy scale parameter $R$, and~the correlation strength $\lambda$ were determined, and~their centrality and transverse mass ($\mT$) dependence was~investigated.

\textls[-5]{The $\alpha$ parameter was found to be centrality-dependent, but~constant in $\mT$, with~the average values ranging between 1.6 and 2.0. A~decreasing trend with $\mT$ and as the collisions become more peripheral was observed for the $R$ parameter, which could be explained by the hydrodynamic-like scaling and the geometrical interpretation, respectively. The~$\lambda$ parameter showed a decreasing trend with $\mT$, but~after removing the effects of the lack of particle identification, a~constant behavior was obtained. A~decrease toward more central collisions was also observed for $\lambda$.}

%%%%%%%%%%%%%%%%%%%%%%%%%%%%%%%%%%%%%%%%%%
\vspace{6pt} 

%%%%%%%%%%%%%%%%%%%%%%%%%%%%%%%%%%%%%%%%%%
%% optional
%\supplementary{The following supporting information can be downloaded at:  \linksupplementary{s1}, Figure S1: title; Table S1: title; Video S1: title.}

%%%%%%%%%%%%%%%%%%%%%%%%%%%%%%%%%%%%%%%%%%

\funding{B. K{\'o}rodi %MDPI: Information regarding the funder and the funding number should be provided. Please check the accuracy of funding data and any other information carefully. A: Everything is good.
 was supported by the ÚNKP-21-2 New National Excellence Program of the Ministry for Innovation and Technology from the source of the National Research, Development and Innovation Fund. This research was supported by the NKFIH OTKA K-138136 and K-128713~grants.}

\dataavailability{The data presented in this study are available on request from the
corresponding author. The data are not publicly available.}
%MDPI: Please provide details regarding where data supporting reported results can be found, including links to publicly archived datasets analyzed or generated during the study. Please refer to suggested Data Availability Statements in section ``MDPI Research Data Policies'' at \url{https://www.mdpi.com/ethics}. You might choose to exclude this statement if the study did not report any data. A: Done.

%\acknowledgments{The author would like to thank the CMS Collaboration.}

\conflictsofinterest{The author declares no conflict of~interest.} 

%%%%%%%%%%%%%%%%%%%%%%%%%%%%%%%%%%%%%%%%%%
%% Optional

%% Only for journal Encyclopedia
%\entrylink{The Link to this entry published on the encyclopedia platform.}

\abbreviations{Abbreviations}{
The following abbreviations are used in this manuscript:\\

\noindent 
\begin{tabular}{@{}ll}
QGP & quark--gluon plasma\\
LHC & Large Hadron Collider\\
HBT & Hanbury Brown and Twiss\\
PbPb & lead--lead\\
CMS & Compact Muon Solenoid\\
AuAu & gold--gold
\end{tabular}
}

%%%%%%%%%%%%%%%%%%%%%%%%%%%%%%%%%%%%%%%%%%
%% Optional
\appendixtitles{yes} % Leave argument "no" if all appendix headings stay EMPTY (then no dot is printed after "Appendix A"). If~the appendix sections contain a heading then change the argument to "yes".
\appendixstart
\appendix
\section[\appendixname~\thesection. Results for Negatively Charged Pairs]{Results for Negatively Charged Pairs} %MDPI: We moved title from \appendixtitles{} to here, please check. A: Good.
\label{a:negativepairs}

\begin{figure}[H]
%\centering
\includegraphics[scale=0.55]{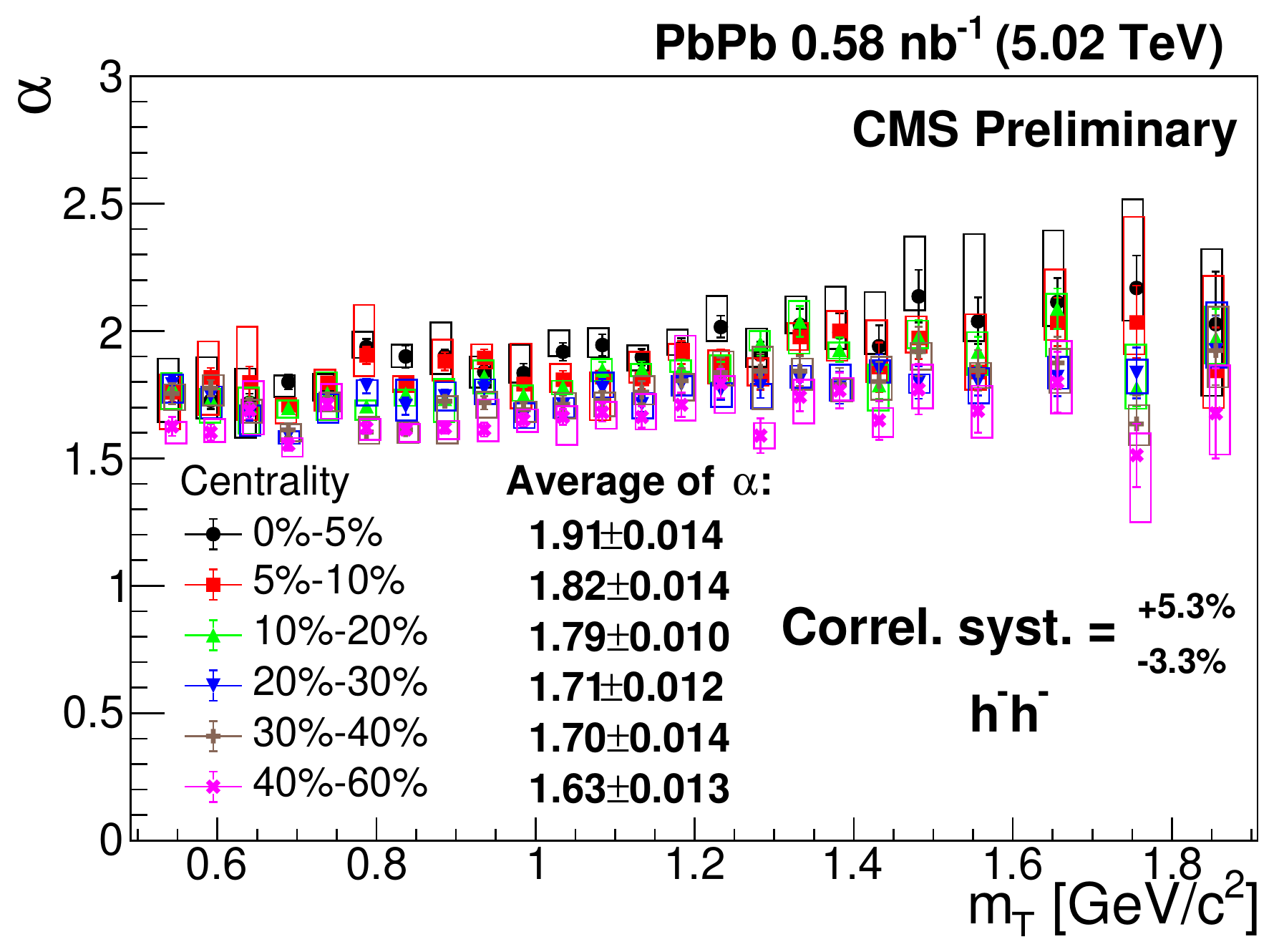}
\caption{The %MDPI: Please change the hyphen (-) into minus sign ($-$, "U+2212"). e.g., "-1" should be "$-$1".
 Lévy stability index $\alpha$ versus the transverse mass $\mT$ in different centrality classes for negatively charged hadron pairs~\cite{PAS}. The~error bars are the statistical uncertainties, while the boxes indicate the uncorrelated systematic uncertainties. The~correlated systematic uncertainty is shown in the~legend.}
\label{f:alpha_neg}
\end{figure}

\begin{figure}[H]
%\centering
\includegraphics[scale=0.55]{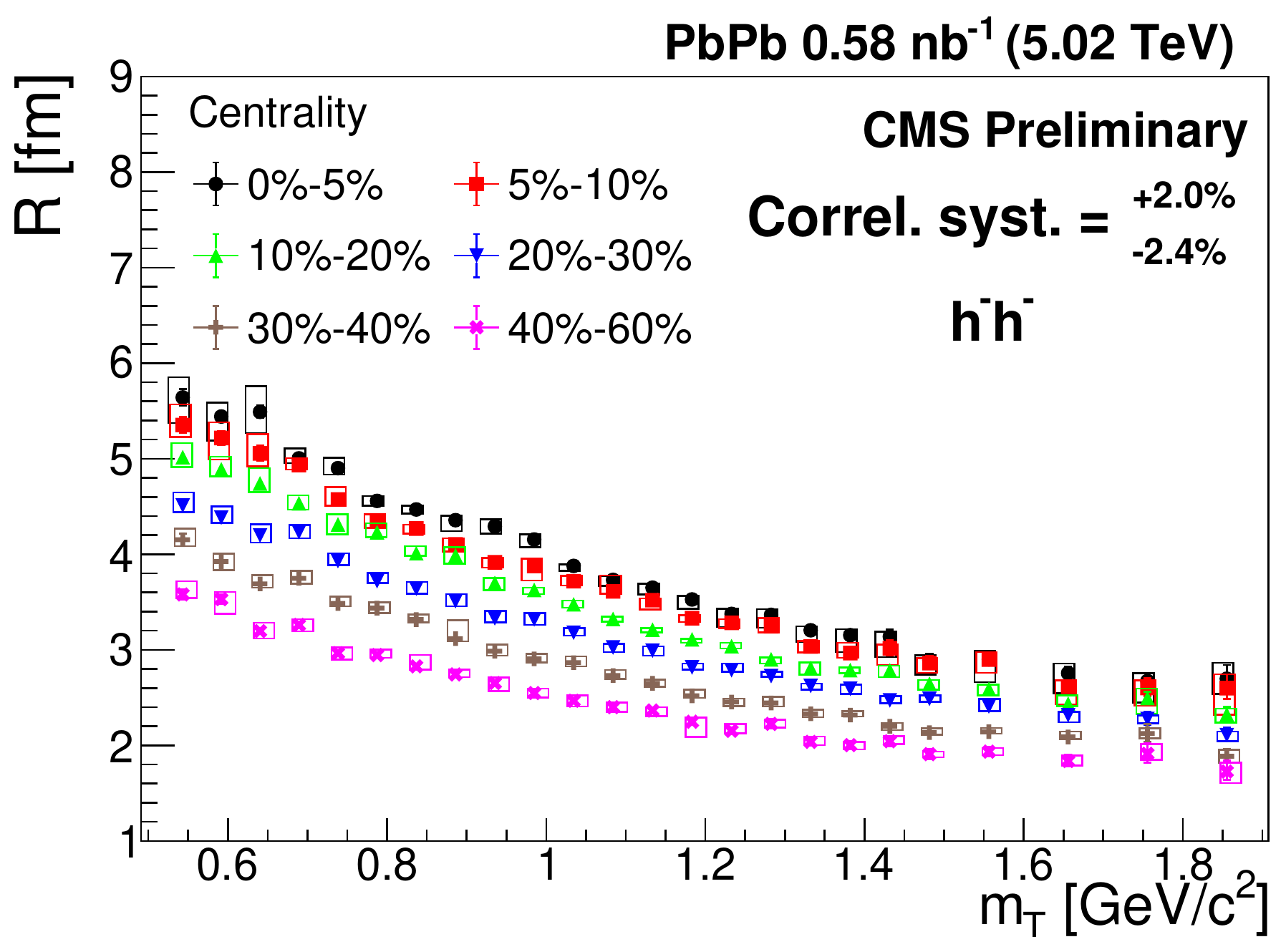}
\caption{The %MDPI: Please change the hyphen (-) into minus sign ($-$, "U+2212"). e.g., "-1" should be "$-$1".
 Lévy scale parameter $R$ versus $\mT$ in different centrality classes for negatively charged hadron pairs~\cite{PAS}. The~error bars are the statistical uncertainties, while the boxes indicate the uncorrelated systematic uncertainties. The~correlated systematic uncertainty is shown in the~legend.}
\label{f:R_neg}
\end{figure}

\begin{figure}[H]
%\centering
\includegraphics[scale=0.55]{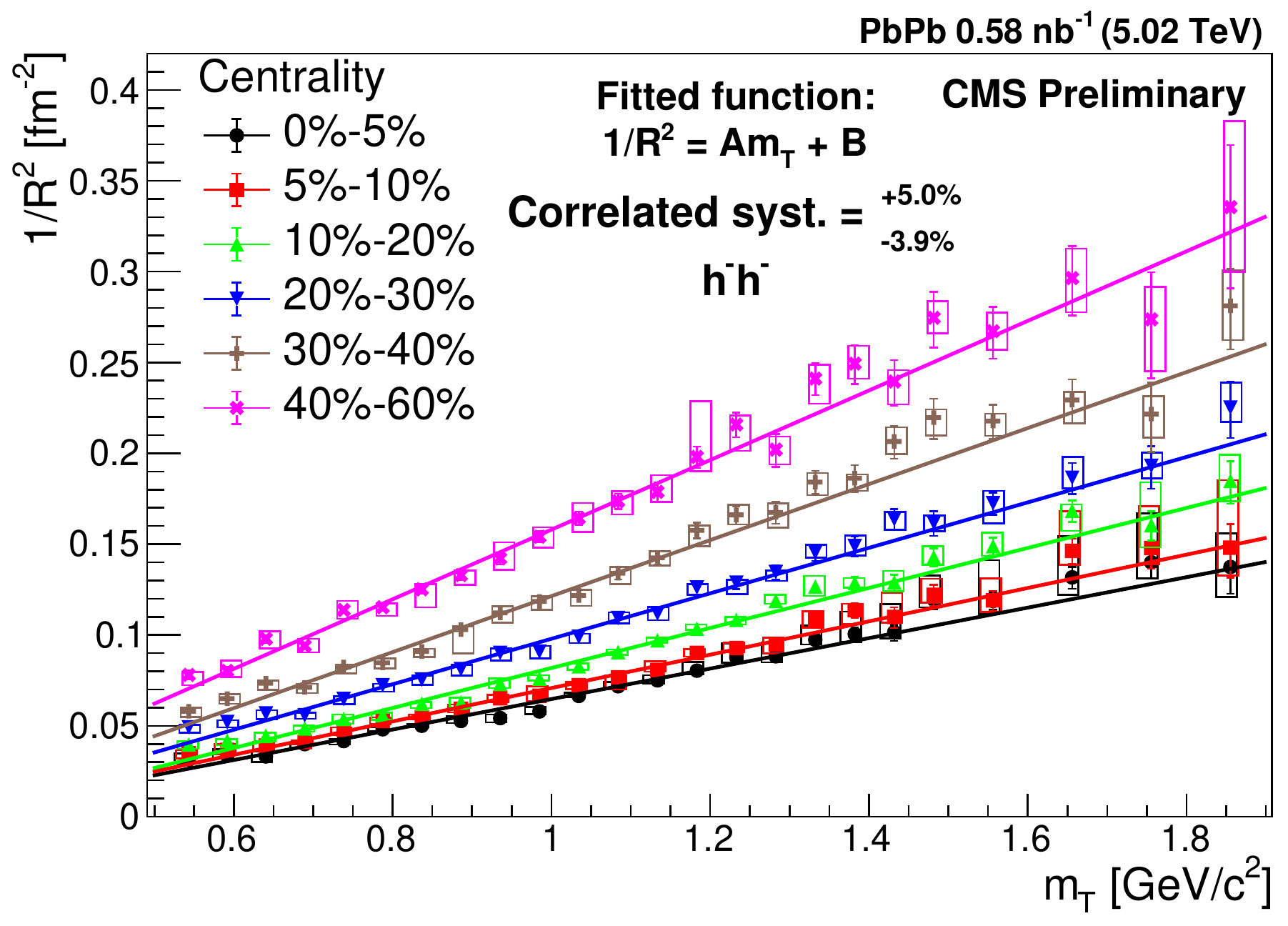}
\caption{The %MDPI: Please change the hyphen (-) into minus sign ($-$, "U+2212"). e.g., "-1" should be "$-$1".
 inverse square of the Lévy scale parameter $R$ versus $\mT$ in different centrality classes for negatively charged hadron pairs~\cite{PAS}. The~error bars are the statistical uncertainties, while the boxes indicate the uncorrelated systematic uncertainties. The~correlated systematic uncertainty is shown in the legend. A~line is fitted to the data for each~centrality.}
\label{f:1_R2_neg}
\end{figure}

\begin{figure}[H]
%\centering
\includegraphics[scale=0.55]{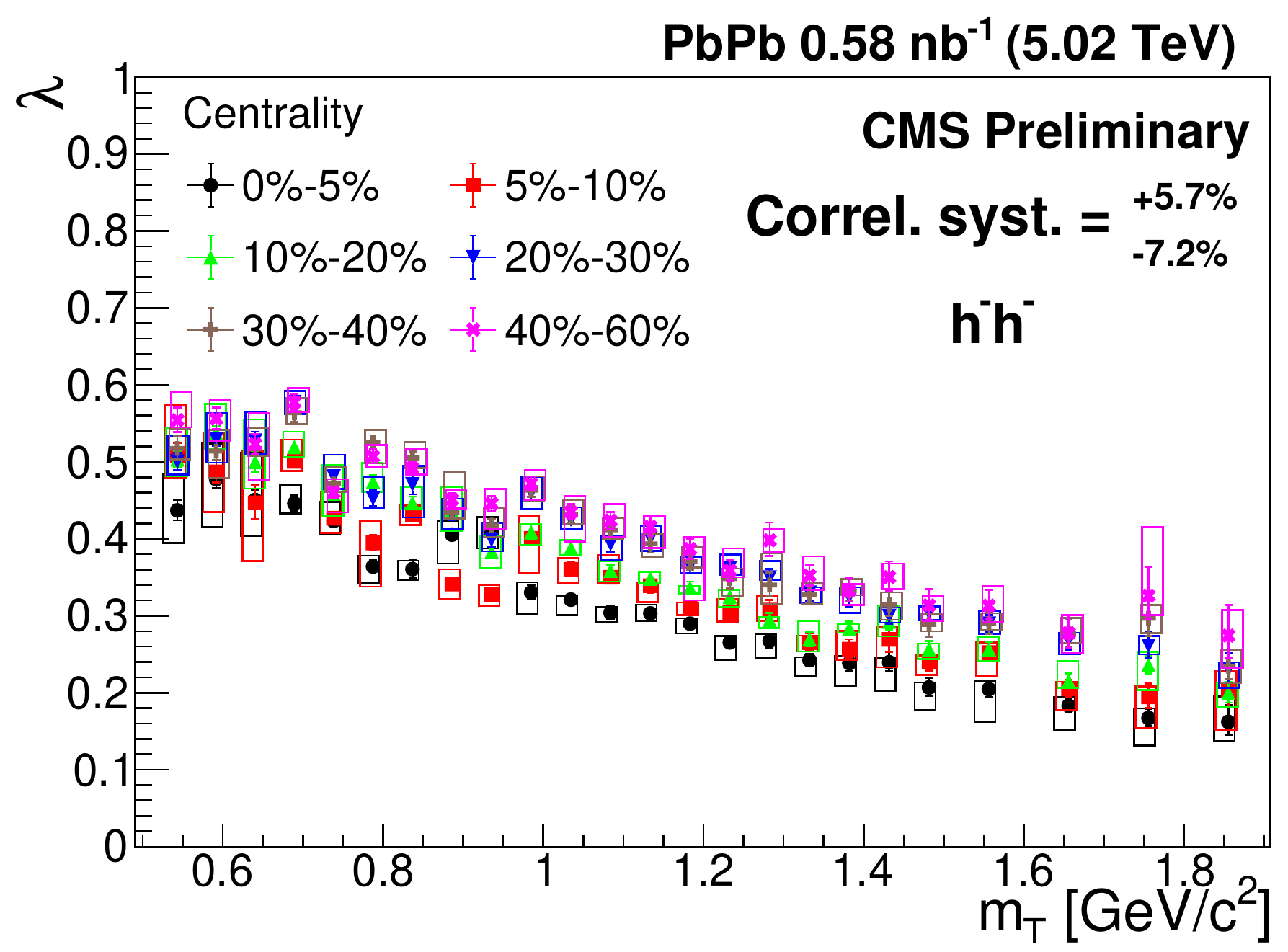}\\
\includegraphics[scale=0.55]{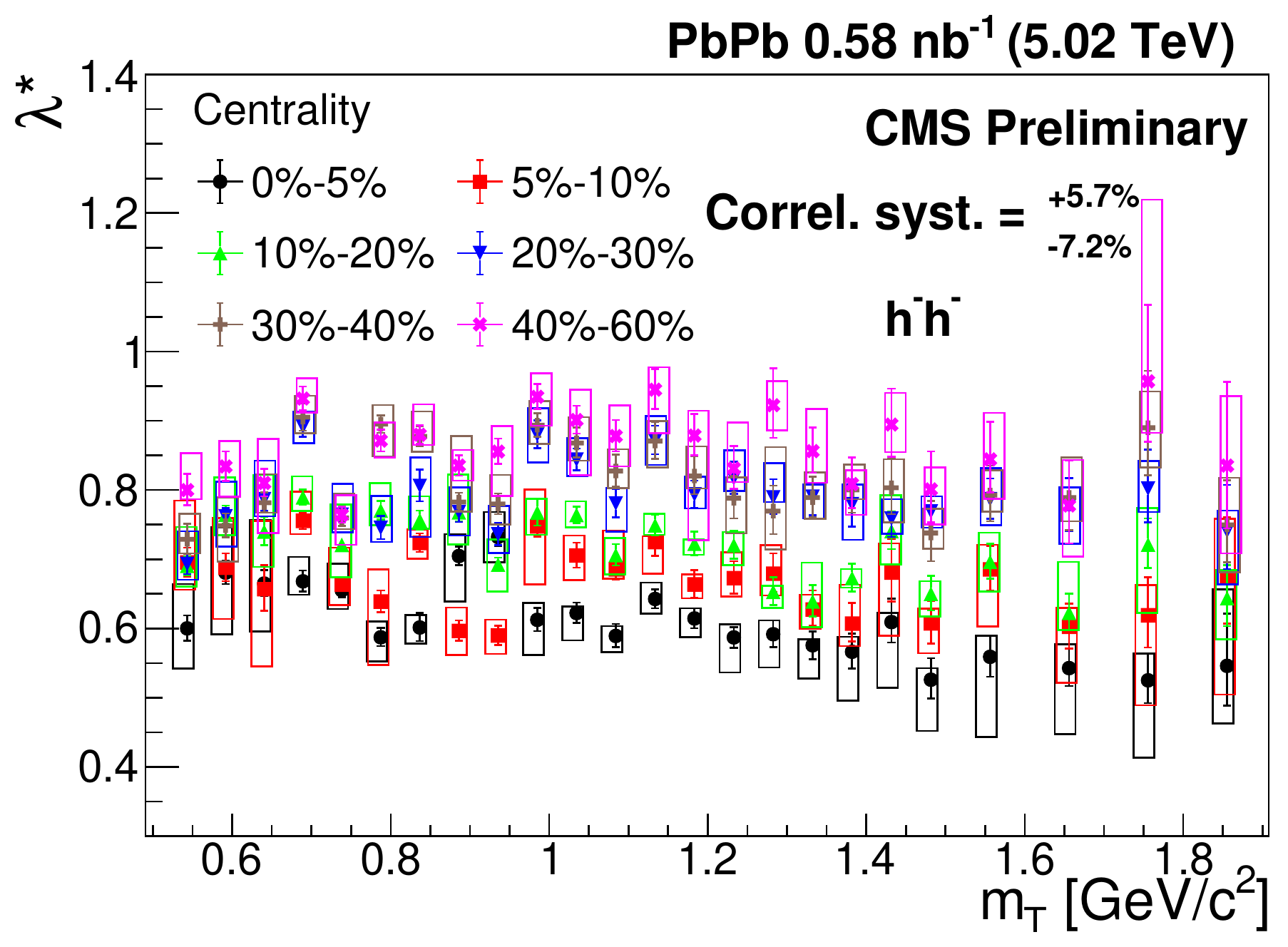}
\caption{The %MDPI: Please change the hyphen (-) into minus sign ($-$, "U+2212"). e.g., "-1" should be "$-$1".
 correlation strength $\lambda$ and the rescaled correlation strength $\lambda^*$ versus $\mT$ in different centrality classes for negatively charged hadron pairs~\cite{PAS}. The~error bars are the statistical uncertainties, while the boxes indicate the uncorrelated systematic uncertainties. The~correlated systematic uncertainty is shown in the~legend.}
\label{f:lambda_neg}
\end{figure}

The results for negatively charged hadron pairs are presented. Due to the fact that they are very similar to the results for positively charged pairs presented in Section~\ref{s:results}, the~interpretations of these results are the same. The~Lévy stability index $\alpha$ is shown as a function of $\mT$ in Figure~\ref{f:alpha_neg}. The~Lévy scale parameter $R$ and its inverse square $1/R^2$ are shown as functions of $m_T$ in Figures~\ref{f:R_neg} and~\ref{f:1_R2_neg}, respectively. The~correlation strength $\lambda$ and the rescaled correlation strength $\lambda^*$ are shown as functions of $m_T$ in Figure~\ref{f:lambda_neg}.

\vspace{-8pt}

\vspace{-8pt}

\vspace{-8pt}

%%%%%%%%%%%%%%%%%%%%%%%%%%%%%%%%%%%%%%%%%%
\begin{adjustwidth}{-\extralength}{0cm}
%\printendnotes[custom] % Un-comment to print a list of endnotes

\reftitle{References}
%=====================================
%\bibliography{sample.bib}

%%%%%%%%%%%%%%%%%%%%%%%%%%%%%%%%%%%%%%%%%%
\PublishersNote{}
\end{adjustwidth}
\end{document}